\documentclass[12pt]{iopart}
\usepackage{graphicx}
\begin{document}

\title[Superconductivity \textit{versus} magnetism in the amorphous palladium “ides”]{Superconductivity \textit{versus} magnetism in the amorphous palladium “ides”: Pd$_{1-c}$(H/D/T)$_{c}$}

\author{I Rodr\'{\i}guez$^1$, R M Valladares$^2$, A Valladares$^2$, D Hinojosa-Romero$^1$ and A A Valladares$^1$}

\address{$^1$Instituto de Investigaciones en Materiales, Universidad Nacional Aut\'{o}noma de M\'{e}xico, Apartado Postal 70-360, Ciudad Universitaria, CDMX, 04510, M\'{e}xico.}
\address{$^2$Facultad de Ciencias, Universidad Nacional Aut\'{o}noma de M\'{e}xico, Apartado Postal 70-542, Ciudad Universitaria, CDMX, 04510, M\'{e}xico.}
\ead{valladar@unam.mx}
\vspace{10pt}
\begin{indented}
\item[]August 2021
\end{indented}

\begin{abstract}
In general, conventional superconductivity and magnetism are competing phenomena. In some alloys this competition is a function of the concentration of the elements. Here we show that in the palladium alloys Pd$_{1-c}$(H/D/T)$_{c}$ (Pd-ides) the increase in the concentration \textit{c} of the ides: hydrogen, deuterium, tritium (H/D/T), lowers the predicted magnetism of amorphous palladium (\textit{a}-Pd) gradually, allowing superconductivity to appear for $ c \approx 40\%$. This magnetism explains why superconductivity does not manifest for smaller values of $c$ ($c \leq 40\%$) in these Pd alloys. Also, these results validate indirectly our predicted magnetism in the amorphous/porous palladium (\textit{a/p}-Pd). The understanding of the interplay between magnetism and superconductivity may contribute to the comprehension of the magnetic behavior in materials, especially in high $T^{*}_{c}$ superconductors, with the corresponding implications.
\end{abstract}

\vspace{2pc}
\submitto{\SUST}
%
 

\section{Introduction}

Conventional superconductivity is a small energy effect as indicated by the low transition temperatures at which it occurs, whereas ferromagnetism, typified by the Curie transition temperature, exists at much higher temperatures. However, in reentrant superconductivity as the temperature is lowered superconductivity sets in at first, and then the magnetic ordering appears, inhibiting the superconducting behavior. For pure amorphous palladium we reported that ferromagnetism appears as a consequence of the amorphicity \cite{rodriguez_emergence_2019}; we now present calculations that indicate that this magnetism interferes with the possible, expected, appearance of superconductivity of this element in bulk, since as the concentration of the ides increases, the magnetism decreases until it vanishes close to $c \approx 40\%$ and superconductivity appears. Also, these results validate indirectly our predicted magnetism in the amorphous/porous palladium (\textit{a/p}-Pd).

Experimentally, 1972 was the year and Skoskiewicz was the author \cite{skoskiewicz_superconductivity_1972}. Something as apparently unassuming as the discovery of superconductivity in the palladium hydrides became a Pandora’s box. It has been said that “this discovery was surprising, since the Pd-H system had been the subject of many studies because of its other interesting properties” \cite{kleiner_superconductivity_2016} although no superconducting properties had been considered. It has also been argued that it was then unusual to find superconductivity in this region of the Periodic Table. Whatever the arguments, the system became the subject of exacerbated interest when Stritzker and Buckel \cite{stritzker_superconductivity_1972} discovered the “reverse isotope effect” (RIE) in the palladium deuterides; that is, the heavier the isotope the higher the superconducting transition temperature, as opposed to the formula proposed by Bardeen, Cooper and Schrieffer, BCS, in their theory \cite{bardeen_theory_1957}. Later, Schirber and Northrup in 1974 corroborated these findings in a systematic study of these two ides \cite{schirber_concentration_1974}. Finally, the trilogy was complete when Schirber \textit{et al.} corroborated RIE for the palladium tritides in 1984 \cite{schirber_superconductivity_1984}.  This made the palladium ides relevant since they did not conform to the normal isotope effect (NIE) propounded by BCS. When the experimental RIE became an accepted fact, possible explanations were put forth that span the spectrum from the optical modes generated by the ides vibrations \cite{papaconstantopoulos_band_1978} to the non-\textit{s} pairing in the formation of the Cooper pairs \cite{fay_possibility_1977}.  Considerations concerning the RIE came and went but the reason for this “abnormal” isotope effect has not been duly clarified. 

To place the experimental discovery of the RIE, and the relevance of the Pd-ides in context, it is opportune to remember that when BCS proposed their theory for conventional superconductors the basic assumptions were the formation of Cooper pairs as a consequence of the attraction between pairs of electrons due to the exchange of phonons, and the coherent motion of these pairs that could sustain permanent electrical currents in a superconductor. The Meissner effect, as a fundamental phenomenon, was also included. In fact, the BCS formula for the superconducting transition temperature  

\begin{equation*}
T^{*}_{c} = 1.13 \left[\frac{\hbar\omega_{D}}{k_{B}}\right]exp\left[ - \frac{1}{N(E_{F})V_{0}} \right]
\end{equation*}

\noindent (where $\hbar$ is Planck’s constant, $\omega_{D}$ is a typical frequency associated to the vibrational modes of the material (usually taken as the Debye frequency), $k_{B}$ is Boltzmann constant, $N(E_{F})$ is the electron density of states at the Fermi level and $V_{0}$ is the Cooper pairing potential) contains the ubiquitous frequencies of the vibrational modes. Since $\omega_{D} \propto m^{-1/2}$ where $m$ is the mass of the ions of the material, then $T^{*}_{c} \propto \omega_{D} \propto m^{-1/2}$. This is the normal isotope effect and indicates that the heavier the ions the lower $T^{*}_{c}$ unlike the experimental results for the Pd-ides.

Another point to be considered is the non-superconducting properties of pure crystalline Pd, when it would seem natural to expect the opposite \cite{bennemann_theory_1973,papaconstantopoulos_superconductivity_1975,pinski_electron-phonon_1978}. Pure palladium in its crystalline form is a puzzling element; in the atomic form, Pd has all the energy levels up to and including the 4\textit{d} completely filled with electrons which would make it an inert-gas-like entity. However, when condensed into a solid with a crystalline fcc structure it becomes a metal with some of the \textit{d}-electrons spilling over to the higher energy \textit{sp}-levels, due to the proximity of the atomic energies of the 4\textit{d} and 5\textit{s}5\textit{p} states. But what implications does this change in properties have for bulk Pd? It has been argued that in the crystalline phase, palladium becomes a strong exchange-enhanced material with a Stoner enhancement factor close to 10 \cite{parks_electron-electron_1969}. In the amorphous state, Pd was found to be magnetic \cite{rodriguez_emergence_2019} and calculations of the solid under negative pressures indicate that it may become magnetic also \cite{moruzzi_magnetism_1989}, as do the experimental results for Pd clusters \cite{sampedro_ferromagnetism_2003} and for Pd platelets \cite{Mendoza_1999}. If magnetism underlies the behavior of Pd, then it is understandable that no superconducting properties are displayed for this material; however, are there other Pd-based materials where this underlying behavior of Pd may manifest and hinder superconductivity? This work indicates that, in fact, the Pd-ides are a good example of this competition between magnetism and superconductivity.

\subsection{Antecedents}

Since our ab initio computer simulations of amorphous Pd have shown that this phase could be magnetic \cite{rodriguez_emergence_2019}, we decided to investigate other circumstances where this finding could contribute to explain experiments, thereby indirectly corroborating our simulational results. Experimentally it has been argued that the ides locate themselves in the interstitial places (tetrahedral or octahedral) in the crystalline structure of palladium. However, such a structure may not be stable since these positions, together with the high mobility of the hydrogen atoms, may foster the diffusion and create defective structures. Since we do not know which defective structures may be created, we decide to generate “bulk” defective structures by amorphizing the ides samples for several concentrations and to study their properties. 

\begin{figure}[h]
\centering
\includegraphics[width=10cm,keepaspectratio]{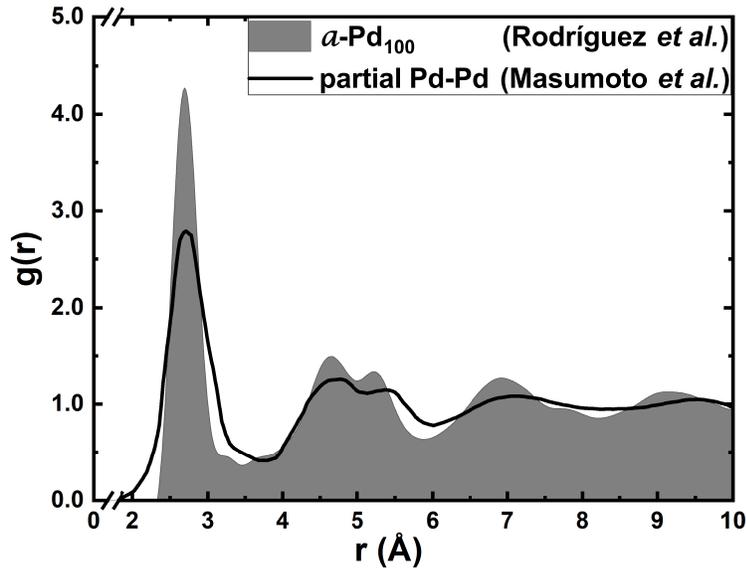}
\caption{\label{fig:fig1} Comparison of the averaged 3 PDFs for our simulated 216-atom supercell of pure amorphous palladium (shaded area) \cite{rodriguez_emergence_2019} with an experimental result due to Masumoto \textit{et al}. for the Pd-Pd partial \cite{masumoto_abstract_1978} (solid black curve) as reported in reference \cite{waseda_structure_1980}.}
\end{figure}

As a reminder, to generate the amorphous samples of pure Pd of Reference \cite{rodriguez_emergence_2019} we first performed molecular dynamics (MD) on a crystalline palladium periodic supercell of 216 atoms with an unstable tetrahedral diamond symmetry to propitiate amorphization, with the experimental density of 12.0 g/cm$^{3}$ and with unrestricted spin. The temperature was maintained constant at 1,500 K using a thermal bath controlled by a Nos\'{e}-Hoover thermostat, during 300 steps of 5 fs each, for a total of 1.5 ps. The cell underwent three independent MD processes followed by geometry optimizations (GO). This amorphizing procedure, a variant of what we have coined as the undermelt-quench process, has been used in our group to generate metallic and semiconducting glassy structures with good results \cite{rodriguez_emergence_2019,wolf_new_2008,romero_new_2010,valladares_new_2011}. Our generated structures for palladium indicate that the three topologies obtained independently are practically indistinguishable as described by their Pair Distribution Functions (PDFs) and are in agreement with the scarce experimental results \cite{masumoto_abstract_1978, waseda_structure_1980}, as depicted in Figure \ref{fig:fig1}.

We propose that when alloying the ides with Pd rather that maintaining the crystalline arrangement of the palladium the material becomes defective; in particular, we propose that it disorders either because the ides locate themselves in the interstitial sites of the lattice mentioned above \cite{yang_formation_2017}, or due to the rearrangement of the material to minimize energy because of the difference in size of the constituents as indicated by the Hume-Rothery rules. In any case, maintaining the basic crystalline structure, although a good first approximation, seems very difficult to sustain since even pure crystalline elements develop vacant sites and interstitial fillings when the temperature of the material is non-zero, a well-known result for many decades \cite{dekker_solid_1957}. So we undertook the task of considering the palladium ides as disordered, and approached the problem by first generating amorphous structures (using three different processes) and then studying their magnetic properties, if any, together with their electronic and topological properties.

There are three interesting questions that have not been satisfactorily answered concerning these Pd-ides: 
\begin{enumerate}
    \item How is it that superconductivity in the palladium hydrides appears for H/Pd ratios close to, but larger than, 0.75 and not for lower ratios? For D the low-limit ratio is 0.71 and for tritides this ratio is close to 0.67, \cite{skoskiewicz_superconductivity_1972,kleiner_superconductivity_2016,stritzker_superconductivity_1972,schirber_concentration_1974,schirber_superconductivity_1984}.
    \item Palladium thin films deposited at very low temperatures are not superconductors; however, when irradiated with low density He$^{+}$ ions superconductivity appears; how does this occur? Superconductivity disappears when this density is increased or decreased. \cite{stritzker_superconductivity_1978,stritzker_superconductivity_1979}.
    \item Is the tetrahedral/octahedral dilemma relevant in the amorphous structures?
\end{enumerate}
In this work we provide answers to those three questions from a new and different angle. For this we shall resort to our above-mentioned recent findings of magnetism in amorphous palladium, obtained through computer simulations. In what follows we shall use the ratio $x = $ides/Pd or the concentration $c = x/(1 + x)$ indistinctively, to refer to the amount of ides in a given sample. 

\section{Method}

The study of magnetism is carried out as in Reference \cite{rodriguez_emergence_2019} and we here outline that procedure, but applied to the Pd-ides supercells. We have now constructed a 256-atom ($4 \times 4 \times 4$ times a face-centered-cubic unit cell) crystalline supercell as the starting point and have generated the (unstable) structures of all the Pd-ides that we studied by randomly substituting palladium with the corresponding ide, H/D/T; a total of 19 concentrations for each of the 3 isotopes. The 19 concentrations varied from 10\% to 60\% of ides ($x = 0.11 - 1.5$) and the interatomic distances were adjusted to make our supercells isodense to the interpolated experimental results \cite{manchester_h-pd_1994}; in this paper we present two prototypical cells only: Pd$_{0.8}$H$_{0.2}$ and Pd$_{0.55}$T$_{0.45}$. To generate the amorphous solid, we performed ab initio NVT MD at 300 K with unrestricted spin, using a thermal bath controlled by a Nos\'{e}-Hoover thermostat, with a time step of 1 fs during 300 steps, for a total duration of 0.3 ps. The code utilized was CASTEP \cite{clark_first_2005} in the Materials Studio (MS) suite of programs \cite{biovia_materials_2016}, and in this spin-unrestricted process we let the spin acquire the value that is congruent with the generated amorphous topology; the fact that we start from an unstable structure fosters the prompt amorphization. Next, we perform the spin-unrestricted GO, using as inputs the magnetism and structures obtained as outputs from the MD processes, to search for the final local energy-minimum amorphous topologies and the corresponding final results for the magnetic moments per atom, all this at $T = 0$. The energy of the final GO structures is indeed minimum and practically constant during the last 10 time-steps for all the converging runs.  We found that the MD runs for ides concentrations higher than 60\% were unstable and did not converge, in agreement with experiment, whereas the structures obtained for concentrations less than or equal to 15\% did not become completely amorphous.

To further investigate the amorphous phase, we did the following. Starting with the amorphous structures generated in Reference \cite{rodriguez_emergence_2019} (216 atoms) we injected only hydrogens (54 and 177 for the structures corresponding to the two concentrations referred to in this paper) to analyze both results for the same ide, making sure that the resulting densities were congruent with the interpolated experimental values and that there was no overlapping among atoms. We then ran MD processes followed by GO processes, and the final results for the PDFs are very similar from the ones reported herein, respectively. This suggests that the final minimum-energy structures are equivalent as characterized by the PDFs. Also, we studied the Plane Angle Distribution functions (PADs) for several triads of atoms, particularly for the Pd-H-Pd trio to investigate the octahedral \textit{vs} tetrahedral dilemma. These results will be discussed in the next section.

To calculate the electronic density of states (eDoS) per atom per spin we obtained the spin-dependent energy levels per energy interval and then smoothed the distribution using a three-point Fast Fourier Transform; the smoothing was the same for all the eDoS to be able to make comparisons. This procedure generated two curves, one for spin up and one for spin down, asymmetrical for $c < 35 \%$ and the difference in the areas under the curves gives the magnetic moment, after multiplication by Bohr's magneton.

\section{Results}

\begin{figure}[h]
\centering
\includegraphics[width=12cm,keepaspectratio]{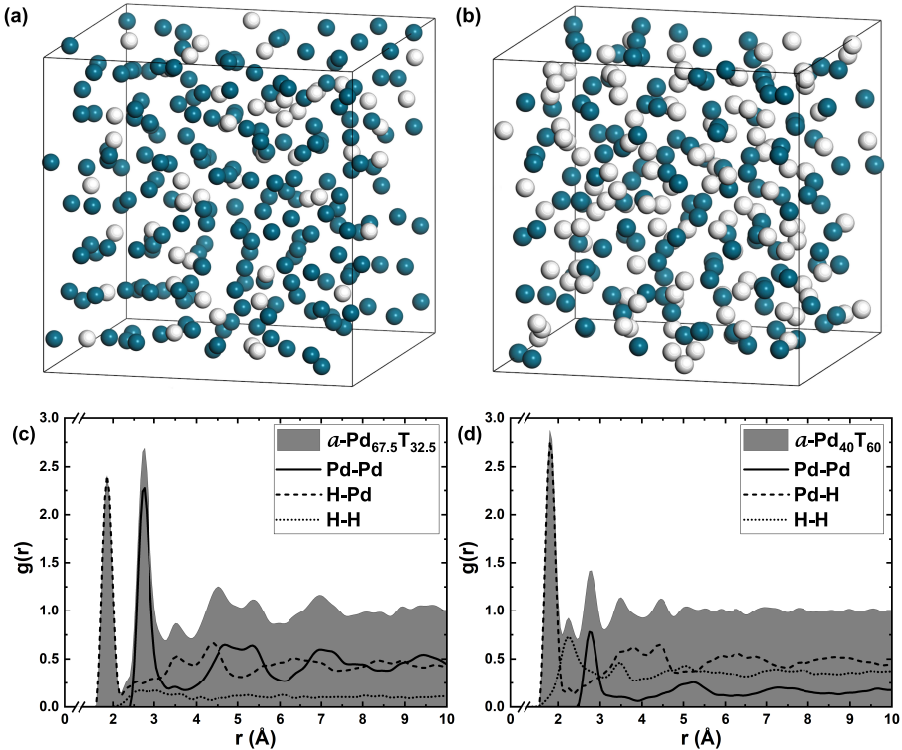}
\caption{\label{fig:fig2} Ball representations of the two 256-atom supercells (Pd, blue (dark) balls; ides, white (light) balls) and PDFs for the two prototypical samples. Atomic structures for (a) an H ratio of 0.25 and (b) a T ratio of 0.82. PDFs for the same samples, (c) an H concentration of 20\% and (d) a T concentration of 45\%.}
\end{figure}

Figures \ref{fig:fig2} (a) and (b) are ball representations of two prototypical supercells corresponding to $x = 0.25$, or $c = 20\%$ of hydrogen, \textit{a}-Pd$_{0.8}$H$_{0.2}$, and $x = 0.82$, or $c = 45 \%$ of tritium, \textit{a}-Pd$_{0.55}$T$_{0.45}$, after applying MD followed by GO starting from the unstable structures. The topologies generated are definitely amorphous, as depicted in Figures \ref{fig:fig2} (c) and (d), where the total and partial smoothed PDFs are displayed. Since the nearest neighbors (nn) distances for the partials H-H and T-T are located at about 2.5 \r{A} this excludes the formation of molecules of ides in the alloys (interatomic molecular distance $\approx 0.74$ \r{A} for an isolated H$_{2}$ molecule \cite{johnson_computational_2002}).

\begin{figure}[h]
\centering
\includegraphics[width=12cm,keepaspectratio]{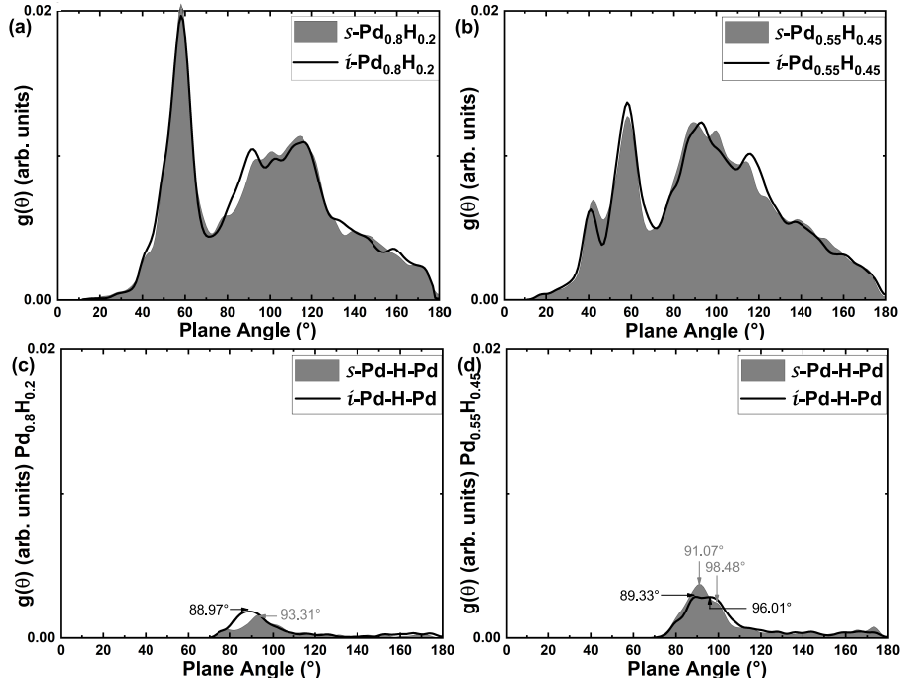}
\caption{\label{fig:fig3} PADs for the two hydride samples with the ide concentrations described in this work. Total PADs, (a) an H ratio of 0.25 and (b) an H ratio of 0.82. Pd-H-Pd partial PADs for the same samples, (c) an H concentration of 20\% and (d) an H concentration of 45\%. The prefixes \textit{i} and \textit{s} are explained in the text.}
\end{figure}

The PADs obtained for these two structures, with hydrogen only to meaningfully compare the results, show an interesting behavior. Here we concentrate on the total PADs and the PADs for the Pd-H-Pd trio as indicated in Figure \ref{fig:fig3}, where we indicate some samples using the prefix \textit{i}, for interstitial, as opposed to those where the substitutional approach was used, identified by the prefix \textit{s}. For the total PADs we notice that as the concentration of H decreases angles at 60° become prominent; this due to the Pd-Pd-Pd trio which becomes more abundant. However, the partial PADs for the trio Pd-H-Pd suggests that the greater the abundance of hydrogen the more conspicuous the angular bimodal tendency becomes, indicating a propensity to the octa/tetra occupation in the amorphous structure. For the lower hydrogen concentrations, there is only one peak at about 90°, suggesting that hydrogens have the tendency to locate themselves in the “octahedral” sites. This answers question iii) asked above.

\begin{figure}[h]
\centering
\includegraphics[width=12cm, keepaspectratio]{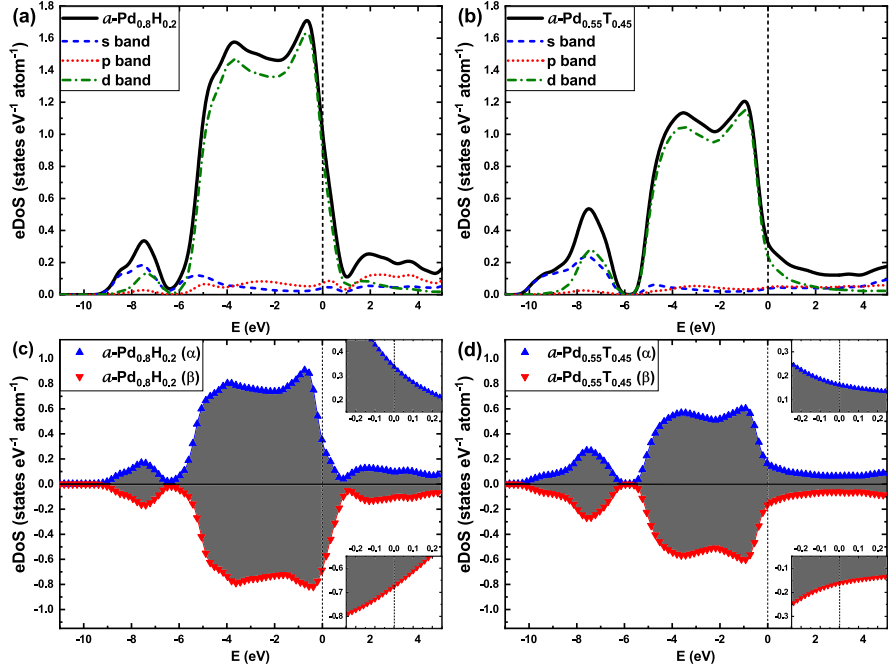}
\caption{\label{fig:fig4} eDoS per atom for the two prototypical supercells. (a) \textit{a}-Pd$_{0.8}$H$_{0.2}$ and (b) \textit{a}-Pd$_{0.55}$T$_{0.45}$ obtained by running both the MD and GO processes with restricted spin (\textit{s}, \textit{p}, \textit{d} and totals are indicated). (c) and (d) depict the spin up and spin down components ran with unrestricted spin; (c) is magnetic and (d) is not. The insets show the behavior in the vicinity of the Fermi energy.}
\end{figure}

Given that the atomic electronic structure of the three ides is the same (except for minuscule effects due to the reduced-mass factor) we would not expect notable changes in the eDoS of the bulk among them, for each concentration. However, due to the difference in the nuclei masses for the three ides, the vibrational density of states (vDoS) does vary drastically as reported in Reference \cite{rodriguez_enhanced_2020}. Figures \ref{fig:fig4} (a) and (b) represent the contributions of the \textit{s}, \textit{p} and \textit{d} states to the total eDoS per atom for the two prototypical supercells: \textit{a}-Pd$_{0.8}$H$_{0.2}$ and \textit{a}-Pd$_{0.55}$T$_{0.45}$ obtained by running both the MD and GO processes with restricted spin. Figures \ref{fig:fig4} (c) and (d) depict the spin up and spin down components of the total eDoS performed with unrestricted spin for the same two supercells. The asymmetry between the up and down curves for \textit{a}-Pd$_{0.8}$H$_{0.2}$ is noticeable, indicating that magnetism is present. The spin up and spin down curves for \textit{a}-Pd$_{0.55}$T$_{0.45}$ are symmetrical, indicating the absence of magnetism. The magnetic results are almost identical for the three ides, for each concentration, as shown in Figure 5.

For concentrations below 35\% the net magnetic moment is different from zero, and diminishes systematically as the concentration increases, until we reach this value. For higher concentrations the magnetic moments become zero for all practical purposes (less than, or of the order of $10^{-4}$ in Bohr magnetons), which indicates that, for these supercells, the substitution of H/D/T for Pd does not generate magnetism, and higher ides concentrations are even unstable. Notice that Skoskiewicz \cite{skoskiewicz_superconductivity_1972} finds superconductivity for $c \approx 45\%$ with a $T^{*}_{c} \geq 1$ K for PdH and that Schirber \textit{et al}. \cite{schirber_superconductivity_1984} find superconductivity for PdT with a $T^{*}_{c} \geq 1$ K at $c \approx 42\%$. 

\begin{figure}[h]
\centering
\includegraphics[width=10cm,keepaspectratio]{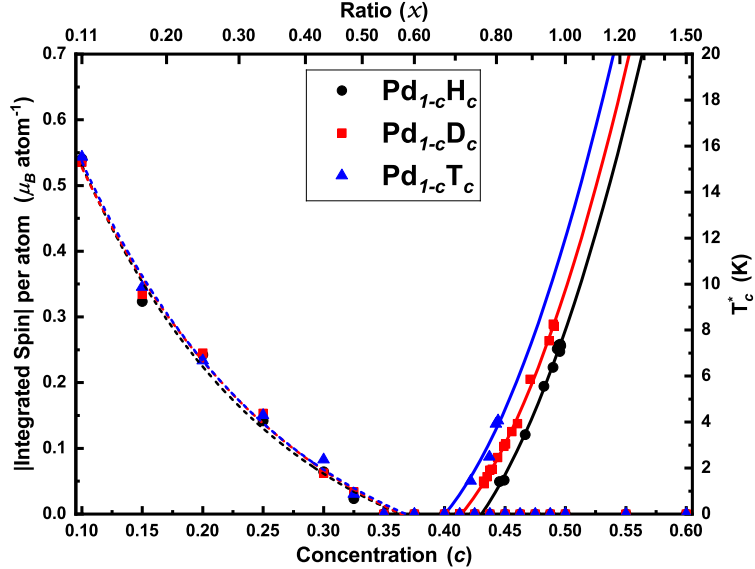}
\caption{\label{fig:fig5} Magnetism (simulated, left vertical axis, dashed curves) and superconducting transition temperatures (experimental, right vertical axis, solid curves), as a function of the concentration \textit{c} (or the ratio \textit{x}) for the H/D/T ides. This behavior is not fortuitous; superconductivity appears after magnetism disappears.}
\end{figure}

In Figure \ref{fig:fig5} we present these conclusions; on the left vertical axis, we have plotted the magnetic moment per atom in units of $\mu_{B}$ for the simulated samples studied (dashed lines), and on the right vertical axis we plot the experimental superconducting transition temperatures for the three ides (solid lines), all function of concentration. We adjusted the experimental points to the same type of quadratic curve $T^{*}_{c} = Ac^{2} + Bc + Intercept$ and the similarities are remarkable; the fitting parameters are $A = 551.4$, $B = -397.76$ and $Intercept = 69.04$ for hydrogen, $A = 577.8$, $B = -414.84$ and $Intercept = 72.69$ for deuterium, and $A = 578.0$, $B = -406.09$ and $Intercept = 68.73$ for tritium; whereas the coefficients of determination (R$^{2}$) are: 0.996, 0.998, 0.914, respectively.

\section{Discussion}

We can see that superconductivity appears once magnetism disappears; this coincidence cannot be fortuitous, it is highly suggestive that for the Pd-ides, superconductivity is inhibited by magnetism. Clearly then, the experimental fact that superconductivity appears for concentrations (ratios) larger than 40\% (0.67) can be understood in terms of this competition between magnetism and superconductivity. This answers question i) posed above.

To look at question ii) we propose to re-interpret the very interesting results by Stritzker \cite{stritzker_superconductivity_1978,stritzker_superconductivity_1979}. Depositing metallic thin films at liquid helium temperatures is the way to generate amorphous, disordered structures; an appealing example is the generation of amorphous bismuth that becomes superconductive \cite{mata-pinzon_superconductivity_2016} at about 6 K, when deposited at temperatures below liquid helium. Palladium should not be an exception and the thin films used by Stritzker in his experiments were most probably amorphous and therefore, according to the results reported in Reference \cite{rodriguez_emergence_2019}, magnetic. We have attributed magnetism to some unbalance of the electronic bonds in amorphous palladium, and this magnetism inhibits its possible superconductivity. Since in the fcc crystalline solid each Pd has 12 nn, the fact that in our \textit{a}-Pd the nn number is smaller than 12 indicates that there might be some “dangling” bonds and some electrons may be loosely bound \cite{rodriguez_emergence_2019}. If a beam of positively charged He ions is directed to the palladium thin films (that are not superconductors because they should be magnetic due to the disorder; although no PDFs have been found in the literature to corroborate this statement) it may be possible that the He$^{+}$ drags some electrons diminishing the number of loose spins in the film, thereby diminishing the magnetic moment per atom in the sample. If magnetism diminishes, superconductivity appears, as found. There must be an optimum density of the He$^{+}$ beam so the dragging process is also optimum: that is why it is found that for low and for high densities the value of the superconducting transition temperature diminishes \cite{stritzker_superconductivity_1978,stritzker_superconductivity_1979}. It would be interesting if these experiments were repeated and the structure of the Pd film was determined; also, it would be desirable if the He$^{+}$ beams were studied after passing the film to see if their ionicity is lost, or at least diminished. In our present work, increasing the concentration of ides in Pd should also increase the neutralization of some of the dangling bonds, decreasing the magnetism of \textit{a}-Pd and of the Pd-ides, thereby fostering the appearance of superconductivity.

\section{Conclusions}

The magnetism found in our computer simulations for amorphous/porous palladium (\textit{a/p}-Pd) is responsible for the delayed appearance of superconductivity in the Pd-ides since the addition of ides neutralizes some of the dangling bonds thereby diminishing their magnetism. This also explains the appearance of superconductivity in palladium thin films deposited at very low temperatures after they are bombarded with He$^{+}$ ions. The dilemma octa/tetra is also discerned by looking at the PADs for PdH with the same concentrations as the prototypical samples reported. The present results indirectly validate our findings for \textit{a}-Pd \cite{rodriguez_emergence_2019} and for \textit{p}-Pd \cite{rodriguez_enhanced_2020}. Finally, if we extrapolate the fitted curves for the experimental $T^{*}_{c}$s, a daring but tempting thing to do, we find the values 222.68 K, 235.65 K and 240.0 K for the transition temperatures of the amorphous 100\% H, D and T ides. 

\ack

I.R.A. thanks DGAPA-UNAM for the postdoctoral fellowship. D.H.R. acknowledges Consejo Nacional de Ciencia y Tecnología (CONACyT) for supporting his graduate studies. A.A.V., R.M.V., and A.V. thank DGAPA-UNAM for continued financial support to carry out research projects under Grants No. IN104617 and IN116520. M.T. V\'{a}zquez and O. Jim\'{e}nez provided the information requested. A. L\'{o}pez and A. Pompa provided technical support and maintenance of the computing unit at IIM-UNAM. Simulations were partially carried out in the Computing Center of DGTIC-UNAM.

\section*{References}
\bibliography{PdH}%

\end{document}